\documentstyle[preprint,aps]{revtex}

\begin{document}
\title{The Local Impurity Self Consistent Approximation (LISA)
to Strongly Correlated Fermion Systems\\
and the Limit of Infinite Dimensions}
\author{{\bf Antoine Georges}}
\address{Laboratoire de Physique Th\'{e}orique de l'Ecole
Normale Sup\'{e}rieure$^1$
\\
24, rue Lhomond; 75231 Paris Cedex 05; France}
\author{{\bf Gabriel Kotliar}}
\address{Serin Physics Laboratory, Rutgers University
\\
Piscataway NJ 08854 USA}
\author{{\bf Werner Krauth} and {\bf Marcelo J. Rozenberg}}
\address{Laboratoire de Physique Statistique de
l'Ecole Normale Sup\'{e}rieure$^2$
\\
24, rue Lhomond; 75231 Paris Cedex 05; France}

\date{October, 16th, 1995}
\maketitle
\begin{abstract}
\noindent
We review the dynamical mean-field theory of strongly correlated
electron systems which is based on a mapping of lattice models onto
quantum impurity models subject to a self-consistency condition.
This mapping is  exact for models of correlated electrons in the limit
of large lattice coordination (or infinite spatial dimensions).
It  extends the standard mean-field
construction from classical statistical mechanics to quantum problems.
We discuss the physical ideas underlying this theory and
its mathematical derivation. Various analytic and numerical
techniques that have been developed recently in order to analyze and
solve the dynamical mean-field equations are reviewed and compared
to each other. The method can be used for the
determination of phase diagrams (by comparing the stability of
various types of long-range order), and the calculation of
thermodynamic properties, one-particle Green's functions,
and response functions.
We review in detail the recent progress in understanding the Hubbard model
and the Mott metal-insulator transition within this approach, including
some comparison to experiments on three-dimensional transition-metal oxides.
We  also  present an overview of the rapidly developing field of
applications of this method to other systems.
The  present limitations of the approach, and possible extensions of the
formalism are finally discussed.
Computer programs for the numerical implementation of this method are
also provided with this article.
\end{abstract}
\smallskip

\medskip
Preprint LPTENS 95/22

\noindent
To appear in Reviews of Modern Physics
\newpage
\contentsline {section}{\numberline {\uppercase {i}}Introduction}{7}
\contentsline {section}{\numberline {\uppercase {ii}}The local impurity
self-consistent approximation: an overview}{15}
\contentsline {subsection}{\numberline {A}Dynamical mean-field equations}{15}
\contentsline {subsection}{\numberline {B}Physical content and connection with
impurity models}{19}
\contentsline {subsection}{\numberline {C}The limit of infinite dimensions}{21}
\contentsline {section}{\numberline {\uppercase {iii}}Derivations of the
dynamical mean-field equations}{24}
\contentsline {subsection}{\numberline {A}The cavity method}{24}
\contentsline {subsection}{\numberline {B}Local nature of perturbation theory
in infinite dimensions}{28}
\contentsline {subsection}{\numberline {C}Derivation based on an expansion
around the atomic limit.}{31}
\contentsline {subsection}{\numberline {D}Effective medium (CPA)
interpretation}{34}
\contentsline {section}{\numberline {\uppercase {iv}}Response functions and
transport}{36}
\contentsline {subsection}{\numberline {A}General formalism}{36}
\contentsline {subsection}{\numberline {B}Frequency-dependent conductivity}{41}
\contentsline {section}{\numberline {\uppercase {v}}Phases with long-range
order}{44}
\contentsline {subsection}{\numberline {A}Ferromagnetic L.R.O}{44}
\contentsline {subsection}{\numberline {B}Antiferromagnetic L.R.O}{45}
\contentsline {subsection}{\numberline {C}Superconductivity and pairing}{47}
\contentsline {section}{\numberline {\uppercase {vi}}Methods of Solution}{50}
\contentsline {subsection}{\numberline {A}Numerical solutions}{51}
\contentsline {subsubsection}{\numberline {1}Quantum Monte Carlo Method}{53}
\contentsline {paragraph}{\numberline {a}Introduction: a heuristic
derivation}{53}
\contentsline {paragraph}{\numberline {b}The Hirsch-Fye algorithm: rigorous
derivation}{55}
\contentsline {paragraph}{\numberline {c}Implementation of the Hirsch-Fye
algorithm}{58}
\contentsline {paragraph}{\numberline {d}The LISA-QMC algorithm and a practical
example}{60}
\contentsline {paragraph}{\numberline {e}relationship with other QMC
algorithms}{62}
\contentsline {subsubsection}{\numberline {2}Exact Diagonalization Method}{64}
\contentsline {subsubsection}{\numberline {3}Comparison of Exact
Diagonalization and Monte Carlo}{71}
\contentsline {subsubsection}{\numberline {4}Spectral densities and real
frequency quantities: Comparison of various methods}{73}
\contentsline {subsubsection}{\numberline {5}Numerical calculation of
susceptibilities and vertex functions}{76}
\contentsline {subsection}{\numberline {B}Analytic methods}{79}
\contentsline {subsubsection}{\numberline {1}Exact methods at low energy}{80}
\contentsline {subsubsection}{\numberline {2}The Iterated Perturbation Theory
Approximation}{83}
\contentsline {subsubsection}{\numberline {3}Slave boson methods and the
non-crossing approximation (NCA)}{86}
\contentsline {paragraph}{\numberline {a}Slave boson mean-field theory.}{87}
\contentsline {paragraph}{\numberline {b}The non-crossing approximation
(NCA).}{89}
\contentsline {subsubsection}{\numberline {4}Equations of motion decoupling
schemes.}{91}
\contentsline {subsection}{\numberline {C}The Projective Self-Consistent
Technique}{92}
\contentsline {section}{\numberline {\uppercase {vii}}The Hubbard model and the
Mott transition}{104}
\contentsline {subsection}{\numberline {A}Early approaches to the Mott
transition}{104}
\contentsline {subsection}{\numberline {B}Models and self-consistent equations
}{106}
\contentsline {subsection}{\numberline {C}Existence of a Mott transition at
half-filling}{109}
\contentsline {subsubsection}{\numberline {1}Metallic phase}{109}
\contentsline {subsubsection}{\numberline {2}Insulating phase}{111}
\contentsline {subsection}{\numberline {D}Phase diagram and
Thermodynamics}{113}
\contentsline {subsubsection}{\numberline {1}Paramagnetic phases}{113}
\contentsline {subsubsection}{\numberline {2}Thermodynamics}{116}
\contentsline {subsubsection}{\numberline {3}Antiferromagnetic phases}{118}
\contentsline {subsection}{\numberline {E}The zero-temperature metal-insulator
transition}{120}
\contentsline {subsection}{\numberline {F}On the $T=0$ instability of the
insulating solution}{123}
\contentsline {subsection}{\numberline {G}Response functions close to the
Mott-Hubbard transition}{126}
\contentsline {subsubsection}{\numberline {1}Magnetic susceptibilities}{126}
\contentsline {subsubsection}{\numberline {2}Charge response and
compressibility}{129}
\contentsline {subsubsection}{\numberline {3}Response to a finite magnetic
field and metamagnetism}{130}
\contentsline {subsection}{\numberline {H}The Hubbard model away from half
filling: doping the Mott insulator}{130}
\contentsline {subsubsection}{\numberline {1}Qualitative arguments}{131}
\contentsline {subsubsection}{\numberline {2}Single particle properties}{132}
\contentsline {subsubsection}{\numberline {3}Thermodynamics}{133}
\contentsline {subsubsection}{\numberline {4}Transport Properties and Response
functions}{135}
\contentsline {subsubsection}{\numberline {5}Phase diagram}{136}
\contentsline {subsection}{\numberline {I}Comparison with Experiments}{137}
\contentsline {subsubsection}{\numberline {1}Phase diagrams}{137}
\contentsline {subsubsection}{\numberline {2}Photoemission spectra}{139}
\contentsline {subsubsection}{\numberline {3}Optical conductivity}{142}
\contentsline {subsubsection}{\numberline {4}Doped Titanates}{145}
\contentsline {section}{\numberline {\uppercase {viii}}Application of the LISA
to various models}{148}
\contentsline {subsection}{\numberline {A}Periodic Anderson Model and the Kondo
lattice}{148}
\contentsline {subsubsection}{\numberline {1}The periodic Anderson model}{148}
\contentsline {subsubsection}{\numberline {2}Half-filled case: Kondo
insulators}{151}
\contentsline {subsubsection}{\numberline {3}The multichannel Kondo
lattice}{155}
\contentsline {subsubsection}{\numberline {4}Metallic quantum
spin-glasses}{157}
\contentsline {subsection}{\numberline {B}The Falicov Kimball Model}{158}
\contentsline {subsection}{\numberline {C}Multiband models: combining LISA and
LDA.}{162}
\contentsline {subsection}{\numberline {D}The extended Hubbard Model and
excitonic effects}{167}
\contentsline {subsection}{\numberline {E}Electron-phonon coupling and the
Holstein model}{173}
\contentsline {subsection}{\numberline {F}Colossal Magnetoresistance and Doped
Manganates}{178}
\contentsline {subsection}{\numberline {G}Systems with quenched disorder}{181}
\contentsline {subsubsection}{\numberline {1}Models of disorder}{181}
\contentsline {subsubsection}{\numberline {2}Interplay of Disorder and SDW or
CDW ordering}{183}
\contentsline {subsubsection}{\numberline {3}Formation of Local Moments and the
Mott Transition in Disordered Systems}{184}
\contentsline {section}{\numberline {\uppercase {ix}}Beyond $d=\infty $:
including spatial fluctuations}{188}
\contentsline {subsection}{\numberline {A}Motivations}{188}
\contentsline {subsection}{\numberline {B}The Bethe-Peierls approximation}{190}
\contentsline {subsection}{\numberline {C}Self-consistent cluster
approximations}{193}
\contentsline {subsection}{\numberline {D} Functional Integral Formulation and
Loop expansion }{197}
\contentsline {section}{\numberline {\uppercase {x}}Conclusion}{202}
\contentsline {section}{\numberline {APPENDIXES\hskip 0pt plus1fill
minus1fill\relax }{}}{204}
\contentsline {section}{\numberline {A}Fermiology in $d=\infty $}{204}
\contentsline {subsection}{\numberline {1}Density of states of some $d=\infty $
lattices}{204}
\contentsline {subsection}{\numberline {2}Momentum dependence of response
functions}{208}
\contentsline {subsection}{\numberline {3}Fermions on the Bethe lattice}{210}
\contentsline {section}{\numberline {B}Details of the Monte Carlo
algorithm}{212}
\contentsline {subsection}{\numberline {1}Some derivations}{212}
\contentsline {subsection}{\numberline {2}Numerical implementation of the QMC
and Gray code enumeration}{213}
\contentsline {subsection}{\numberline {3}Numerical implementation of the
self-consistency condition}{215}
\contentsline {section}{\numberline {C}Details of the Exact Diagonalization
algorithm}{216}
\contentsline {section}{\numberline {D}Access to FORTRAN programs}{219}
\newpage
\begin{verbatim}
----------------------------------------------------------
HOW TO GET ARTICLE by anonymous ftp
----------------------------------------------------------
Suppose your username is  username@usernode.univ.edu. To log on
via anonymous ftp, you should proceed as follows:

ftp ftp.lps.ens.fr
Username: anonymous
Password: username@usernode.univ.edu
          cd  pub/users/lisa

To obtain the whole file (excluding figures), simply type  'get paper.ps'
To obtain the chapters separately, use the following commands:
==================================================================
To obtain:                       |    type:
==================================================================
Abstract & Table of Contents     |    get abstract_toc.ps
chapter 1                        |    get chapter1.ps
chapter 2 - 10                   |    (as above)
Appendices                       |    get appendices.ps
References & figure captions     |    get references_captions.ps
Tables                           |    get tables.ps

There are 87 figures, which are all available in postscript format.
We have stored them as compressed tar archives.
==================================================================
To obtain:                     | type:                  | size
==================================================================
                         first type 'binary'
Figures  1 - 22 (chapter 1 - 6)| get figures1-22.tar.Z  | 3.9 MB
Figures 23 - 64 (chapter 7)    | get figures23-64.tar.Z | 9.8 MB
Figures 65 - 87 (chapters 8 and
    onwards)                   | get figures65-87.tar.Z |12.6 MB

A step-by-step procedure to obtain the complete article including
figures is thus as follows:

get paper.ps    ( or chapter-by-chapter, as explained above)
binary
get figures1-22.tar.Z
get figures23-64.tar.Z
get figures65-87.tar.Z
quit
===============================================================

THEN, on YOUR machine, the .ps files can be readily printed.
In order to produce  printable figure files,  type:

uncompress figures1-22.tar.Z
tar -xvf figures1-22.tar

This produces figures 1 through 22, via files named fig01.final etc...

repeat with the two other figure files.

Some of the figures are scanned hard-copy figures, and are quite
large. Make sure to provide sufficient space on your disk before
loading the whole material. Also the printing of the paper and the
figures may take quite a while, depending on the printer.
&&&&&&&&&&&&&&&&&&&&&&&&&&&&&&&&&&&&&&&&&&&&&&&&&&&&&&&&&&&&&&&&&
All the material is accessible by WWW.
Point at  http://www.lps.ens.fr/~krauth

For problems with the depository (ftp, WWW)  send  mail to
krauth@physique.ens.fr
------------------------------------------------------------------
\end{verbatim}
\end{document}